\def\lae{\mathrel{<\kern-1.0em\lower0.9ex\hbox{$\sim$}}}
\def\gae{\mathrel{>\kern-1.0em\lower0.9ex\hbox{$\sim$}}}

\newcommand{\be}{\begin{equation}}
\newcommand{\ee}{\end{equation}}

\documentclass[apj]{emulateapj}
\usepackage{}
\usepackage{txfonts}
\usepackage{cases}

\shorttitle{Rapid TeV Flaring in Markarian 501} \shortauthors{Zheng
\& Zhang}

\begin{document}

\title{Rapid TeV flaring in markarian 501}
\author{Y. G. Zheng\altaffilmark{1,2} and L. Zhang\altaffilmark{1}}
\altaffiltext{1}{Department of Physics, Yunnan University, Kunming, 650091, China (E-mail:lizhang@ynu.edu.cn)}
\altaffiltext{2}{Department of Physics, Yunnan Normal University, Kunming, 650092, China }

\begin{abstract}
We investigate rapid TeV flaring in markarian 501 in the frame of a
time-dependent one-zone synchrotron self-Compton (SSC) model. In
this model, electrons are accelerated to extra-relativistic energy
through the stochastic particle acceleration and evolve with the
time, nonthermal photons are produced by both synchrotron and
inverse Comtpon scattering off synchrotron photons. Moreover,
nonthermal photons during a pre-flare are produced by the
relativistic electrons in the steady state and those during a flare
are produced by the electrons whose injection rate is changed at
some time interval. We apply the model to the rapid flare of
Markarian 501 on July 9, 2005 and obtain the multi-wavelength
spectra during the pre-flare and during the flare. Our results show
that the time-dependent properties of flares can be reproduced by
varying the injection rate of electrons and  a clear canonical
anti-clockwise-loop can be given.
\end{abstract}


\section{Introduction}
\label{sec:intro}

Variability, which is found from radio to TeV $\gamma$-ray bands, is
one of the major characteristics of blazars. The variability
timescales from a few minutes to days in the optical band have
been extensively investigated (e.g. Sillanp$\ddot{a}\ddot{a}$ et al.
1991; Wagner $\&$ Witzel 1996; Lainela et al. 1999). Particularly in
the X-ray and TeV regimes in which photons are produced by radiation
of ultra-relativistic electrons close to their maximum energy, the
observations of variability timescales constrain on particle
acceleration mechanism in TeV blazars. For examples, Kataoka et al.
(2001) reported the observations of the X-ray flares with timescales
of from hours to days for three TeV blazars (Markarian 421, Markarian 501, and PKS
2155-304), Albert et al. (2007) obtained a rapid TeV variability
with of several ten minutes for Markarian 501 by MAGIC. Observed short
timescales indicate that the variability is associated with small
regions in the relativistic jet, which is located on a distance in
excess of one hundred Schwarzschild radii ($r_{\rm s}$) with a central
black hole mass $M=10^{9}M_{\odot}$, rather than the center region
(Begelman et al. 2008). Relativistic particles maybe responsible for
the emission flare. These particles are ejected from the central
region alone with the subsistent jet structure, and radiate away
their energy at $100r_{\rm s}$ quickly, or the particles are
accelerated within the jet, close to the emission region.

Generally, soft lags can be interpreted as due to electron cooling
(Kirk et al. 1998; Kirk $\&$ Mastichiadis 1999; Kusunose et al.
2000). However, with the fast TeV $\gamma$-ray flare in Markain 501
on July 9, 2005,  the evidence that hard $\gamma$-ray lagged the
soft ones by $4\pm1$ minutes was discovered (Albert et al. 2007).
For explaining these abnormal phenomenon, Bednarek $\&$ Wagner
(2008) proposed the radiating blob accelerating during the flare,
but in their model the particles would only undergo cooling
processes without any accelerating around the high blob Lorentz factor plasma
flow. Mastichiadis $\&$ Moraitis (2008) showed that allowing the
particles to accelerate gradually can explain the observed feature,
and reach the  particles energy to $\gamma\sim10^{6}$, the
acceleration timescales is of the order of hours. Following their
model, Tammi $\&$ Duffy (2009) compared four different acceleration
mechanisms, and pointed out that the timescale may be too long for
first-order Fermi acceleration, so the stochastic acceleration may
be as a promising candidate for the energy dependent time delays.

Motivated by above arguments, we study the time-dependent one-zone
synchrotron self-Compton (SSC) model in the presence of stochastic
particle acceleration, and then apply the model to Markarian 501 for
explaining its flare and time delay properties, especially the rapid
flare of Markarian 501 on July 9, 2005. Throughout the paper, we assume
the Hubble constant $H_{0}=70$ km s$^{-1}$ Mpc$^{-1}$, the matter
energy density $\Omega_{\rm M}=0.27$, the radiation energy
density$\Omega_{\rm r}=0$, and the dimensionless cosmological
constant $\Omega_{\Lambda}=0.73$.

\section{The Model}
\label{sec:model}

Assuming the accelerated particles have an isotropic diffusion in
momentum space, the evolution of the energetic particle distribution
can be described by the momentum diffusion equation (Tverskoi 1967):
\begin{equation}
\frac{\partial f(p,t)}{\partial t}=\frac{1}{p^{2}}\frac{\partial
}{\partial p}[p^{2}D(p,t)\frac{\partial f(p,t)}{\partial p}]\;,
\label{Eq1}
\end{equation}
where $f(p,t)$ is the isotropic, homogeneous phase space density,
$p$ the dimensionless particle momentum, $p=\beta\gamma$, $D(p,t)$
the momentum-diffusion coefficient due to interactions with
magnetohydrodynamic waves, $\gamma$ the particle Lorentz factor, and
$\beta$ the particle velocity in units of light velocity $c$. The
particle number density $N(p,t)=4\pi p^2f(p, t)$ is directly related
to the phase space density.

For a specific source, after including injection, radiation, and
escape of the particles, Eq. (\ref{Eq1}) can be rewritten as (Katarzynski et
al. 2006)
\begin{eqnarray}
\frac{\partial N(\gamma,t)}{\partial t}&=&\frac{\partial}{\partial
\gamma}\{[C(\gamma,t)-A(\gamma,t)]N(\gamma,t)+D(\gamma,t)\frac{\partial
N(\gamma,t)}{\partial
\gamma}\}\nonumber\\&+&Q(\gamma,t)-E(\gamma,t)\;,
\label{Eq2}
\end{eqnarray}
where we have assumed that the particles are ultra-relativistic,
$\beta\approx 1$, so the momentum becomes equivalent to the Lorentz
factor of particle ($p=\gamma$). In Eq. (\ref{Eq2}),
$C(\gamma,t)=(d\gamma/dt)_{\rm syn} + (d\gamma/dt)_{\rm IC}$ is the radiative cooling parameter that
describes the synchrotron and inverse-Compton cooling of the
particles at time $t$. For the synchrotron cooling, $(d\gamma/dt)_{\rm syn}=(4/3)(\sigma_{\rm T}c/m_{\rm e}c^2)U_{\rm B}(t)\gamma^2$
is the rate of the synchrotron loss, $U_{\rm B}$ is the energy densities of the  magnetic field, $m_{\rm
e}$ is the electron rest mass, and $\sigma_{\rm T}$ is the Thomson
cross section. For the IC cooling, the Klein-Nishina (KN) effects at high energy will be important and will modify the electron distribution and
the inverse Compton spectrum (e.g., Moderski et al. 2005; Nakar et al. 2009), The rate of inverse Compton energy losses in which
the KN corrections is included is given by (Moderski et al. 2005):
\begin{equation}
(\frac{d\gamma}{dt})_{\rm IC}=\frac{4\sigma_{T}c}{3m_{e}c^{2}}U_{\rm rad}(\gamma, t)\gamma^{2}F_{\rm KN}\;,
\label{Eq3}
\end{equation}
where, $U_{\rm rad}(\gamma,t)=\int_{\epsilon_{\rm 0,min}}^{\epsilon_{\rm 0,max}}U(\epsilon_{\rm 0})d\epsilon_{\rm 0}$ is the total energy density of the radiation field,
$U(\epsilon_{\rm 0})$ is the energy distribution of the soft photons, $\epsilon_{\rm 0}$ is the soft photons energy of the synchrotron radiation,
$F_{\rm KN}=[1/U_{\rm rad}(\gamma,t)]\int_{\epsilon_{\rm 0,min}}^{\epsilon_{\rm 0,max}}f_{\rm KN}(\chiup)U(\epsilon_{\rm 0})d\epsilon_{\rm 0}$,
$\chiup=4\gamma\epsilon_{\rm 0}$, the function $f_{\rm KN}$ can be approximated as (Moderski et al. 2005) as follow:
\begin{equation}
f_{\rm KN}\simeq\left\{ \begin{array}{ll}
1          & ~\chiup\ll 1 ~\mbox{(Thomson limit)}\\
\frac{9}{2\chiup^2}(\ln\chiup-\frac{11}{6})     & ~ \chiup\gg 1~ \mbox{(KN limit)}\;.
\end{array} \right.
\label{Eq4}
\end{equation}
When $\chiup\lesssim  10^{4}$, $f_{\rm KN}\simeq 1/(1+\chiup)^{3/2}$. Therefore, the  radiative cooling parameter is given by
\begin{equation}
C(\gamma,t)=\frac{4}{3}\frac{\sigma_{\rm T}c}{m_{\rm e}c^2}[U_{\rm B}(t)+U_{\rm
rad}(\gamma, t)F_{\rm KN}]\gamma^2\;.
\label{Eq5}
\end{equation}

Other terms in the left side of Eq. (\ref{Eq2}) are as follows: $A(\gamma, t)=\gamma/t_{\rm acc}$ is the acceleration
term that describes the particle energy gain per unit time, which is given by
\begin{equation}
A(\gamma, t)=\frac{\gamma}{t_{\rm acc}}=\frac{2D(\gamma, t)}{\gamma}\;,
\label{Eq6}
\end{equation}
where the acceleration time $t_{\rm acc}=\gamma^2/2D(\gamma, t)$ is used;
$E(\gamma,t)$ represents escape term, which is
\begin{equation}
E(\gamma,t)=\frac{N(\gamma, t)}{t_{\rm esc}}=\frac{c}{R}N(\gamma, t)\;,
\label{Eq7}
\end{equation}
where
escape timescale $t_{\rm esc}=R/c$ depends on the the emission
region size $R$; $Q(\gamma,t)$ is the source term, here we consider
continuous injection case, i.e. the particles are continuously
injected at the lower energy ($1\le \gamma\le2$) and systematically
accelerated up to the equilibrium energy ($\gamma_{\rm e}$),
where the acceleration process is fully compensated for by the cooling,
i.e. $t_{\rm cool}(\gamma_{\rm e})=t_{\rm acc}$.

In time-dependent one-zone SSC model, Eq. (\ref{Eq2}) needs to be solved by
a numerical method because of the non-linearity process involved. We
adopt an implicit difference scheme given by Chang $\&$ Copper
(1970). In our calculations, we adopt the forward
differentiation in time and the centered differentiation in the
energy (see Press et al. (1989) for a detail discussion). The
merits of the implicit difference scheme are as follows: 1) solution
is always positive; 2) particle number is always conserved; and 3)
we can significantly reduces the number of mesh points in
calculation with no loss of accuracy. We define the energy mesh
points of electrons with logarithmic steps:
\begin{equation}
\gamma_{\rm j}=\gamma_{\rm min}(\frac{\gamma_{\rm inf}}{\gamma_{\rm min}})^{\frac{(j-1)}{(j_{\rm max}-1)}};j=1,2,3,...,j_{\rm max}\;,
\label{Eq8}
\end{equation}
where $j_{\rm max}$ is the number of the mesh points, $\gamma_{\rm
min}$ and $\gamma_{\rm inf}$ are the minimum and maximum Lorentz
factors of electrons to be used in the calculation, respectively. In
our calculation, a grid of 200 points has been used both for
particles energy and photon frequency. Since we assume an
exponential cut-off at $\gamma=\gamma_{\rm max}$ for the particles
distribution $N(\gamma,t)$, $\gamma_{\rm inf}$ is taken to be much
larger than $\gamma_{\rm max}$, here we adopt $\gamma_{\rm
inf}=10^9$. By defining $r=(\gamma_{\rm inf}/\gamma_{\rm
min})^{(1/j_{\rm max})}$, the energy intervals can be expressed as
$\Delta\gamma_{\rm j}=(r-1)\gamma_{\rm j}$, $\Delta\gamma_{\rm
j+1/2}=(r-1)\gamma_{\rm j+1/2}=(1/2)(r-1)(\gamma_{\rm
j+1}+\gamma_{\rm j})$, $\Delta\gamma_{\rm j-1/2}=(r-1)\gamma_{\rm
j-1/2}=(1/2)(r-1)(\gamma_{\rm j}+\gamma_{\rm j-1})$ (e.g., see Park
\& Petrosian 1996). Quantities with the subscript $j\pm1/2$ are
calculated at half grid points. In order to discretize the
continuity equation, we define
\begin{equation}
G(\gamma,t)=[C(\gamma,t)-A(\gamma,t)]N(\gamma,t)+D(\gamma,t)\frac{\partial
N(\gamma,t)}{\partial \gamma}
\label{Eq9}
\end{equation}
and $N^{n}_{\rm j}=N(\gamma_{\rm
j},n\Delta t)$. Therefore, Eq. (\ref{Eq2}) can be written as
\begin{equation}
\frac{N_{\rm j}^{n+1}-N_{\rm j}^{n}}{\Delta t}=\frac{G_{\rm j+1/2}^{n+1}-G_{\rm j-1/2}^{n+1}}{\Delta\gamma_{\rm j}}+Q_{\rm j}^{n}-\frac{N_{\rm j}^{n+1}}{t_{\rm esc}}\;,
\label{Eq10}
\end{equation}
where
\begin{equation}
G_{\rm j+1/2}^{n+1}=[C_{\rm j+1/2}^{n+1}-A_{\rm
j+1/2}^{n+1}]N_{\rm j+1/2}^{n+1}+D_{\rm j+1/2}^{n+1}\frac{ N_{\rm
j+1}^{n+1}-N_{\rm j}^{n+1}}{ \Delta\gamma_{\rm j+1/2}}\;,
\label{Eq11}
\end{equation}
\begin{equation}
G_{\rm
j-1/2}^{n+1}=[C_{\rm j-1/2}^{n+1}-A_{\rm j-1/2}^{n+1}]N_{\rm
j-1/2}^{n+1}+D_{\rm j-1/2}^{n+1}\frac{ N_{\rm j}^{n+1}-N_{\rm
j-1}^{n+1}}{ \Delta\gamma_{\rm j-1/2}}\;.
\label{Eq12}
\end{equation}
In this case, we have
$N_{\rm j+1/2}=(1/2)(N_{\rm j+1}+N_{\rm j})$, $A_{\rm
j+1/2}=(1/2)(A_{\rm j+1}+A_{\rm j})$, $C_{\rm j+1/2}=(1/2)(C_{\rm
j+1}+C_{\rm j})$, $D_{\rm j+1/2}=(1/2)(D_{\rm j+1}+D_{\rm j})$, and
$N_{\rm j-1/2}=(1/2)(N_{\rm j}+N_{\rm j-1})$, $A_{\rm
j-1/2}=(1/2)(A_{\rm j}+A_{\rm j-1})$, $C_{\rm j-1/2}=(1/2)(C_{\rm
j}+C_{\rm j-1})$, $D_{\rm j-1/2}=(1/2)(D_{\rm j}+D_{\rm j-1})$. With
the energy interval $\Delta\gamma$ and time interval $\Delta t$,
using the no-flux boundary condition (Park \& Petrosian 1995), Eq.
(2) can be written in a tri-diagonal matrix and can be solved by
numerical approach (e.g. Press et al. 1989). If the electron number
density $N(\gamma_{j},0)$ at time $t=0$ is given, then the number
density $N(\gamma_{j},\Delta t)$ can be calculated at time $t=\Delta
t$. The iteration of above prescription gives the electron number
density at an arbitrary time $t$ (e.g., Chaiberge $\&$ Ghisellini
1999).

After calculating the
electron number density $N(\gamma, t)$ at a time $t$, we can use the
formulae given by Katarzynski et al. (2001) to calculate the
synchrotron intensity $I_{\rm s}(\nu, t)$ and the intensity of
self-Compton radiation $I_{\rm c}(\nu, t)$ , and then calculate the
flux density observed at the Earth as follows:
\begin{equation}
F_{\rm tot}(\nu, t)=\pi\frac{R^2}{d^2}\delta^3(1+z)[I_{\rm s}(\nu,
t)+ I_{\rm c}(\nu, t)]\;,
\label{Eq13}
\end{equation}
where $d$ is the luminosity distance, $z$ is the redshift, and
$\delta = [\Gamma(1 - \beta\cos \theta)]^{-1}$ is the Doppler factor
where $\Gamma$ is the blob Lorentz factor, $\theta$ is the angle of
the blob vector velocity to the line of sight and $\beta= v/c$.
Since at high energies the Compton photons may produce pairs by
interacting with the synchrotron photons, this process may be
decrease the observed high energy radiation (Coppi \& Blandford
1990; Finke et al. 2009). Katarzynski et al. (2001) analyze the
absorption effect due to pair-production inside the source, they
found that its process is almost negligible. On the other hand, very
high energy (VHE) photons from the source are attenuated by
photons from the extragalactic background light (EBL). Therefore,
after taking the absorption effect, the flux density observed at the
Earth becomes
\begin{equation}
F(\nu)=F_{\rm tot}(\nu, t)\exp[-\tau(\nu,z)]\;,
\label{Eq14}
\end{equation}
where $\tau(\nu,z)$ is the absorption optical depth due to
interactions with the EBL (Kneiske et al. 2004; Dwek \& Krennrich
2005). In our calculation, we use the absorption optical depth which
is deduced by the average EBL model in Dwek \& Krennrich (2005).

\section{Validation of the Numerical Code}

In order to validate our numerical code, we compare the time
evolution of the electron spectrum calculated in our code with the
analytic solutions given by Chang \& Cooper (1970). In our
calculation, we assume that electrons lose energy by synchrotron and
IC cooling, where the IC cooling is
assumed to occur in the Thomson regime. Since the loss rates of both
synchrotron radiation and inverse Compton scattering satisfy
$d\gamma/dt\propto \gamma^{2}$, we can write the characteristic
cooling time as $t_{\rm cool}(\gamma)=1/C_{0}\gamma$ with a cooling
coefficient $C_{0}$. Otherwise, the system has not injection and
escape of the particle during the evolution process (i.e.
$Q(\gamma,t)=0$ and $E(\gamma,t)=0$). Under the above assumptions,
equation (2) can be written as in the steady state ($\partial
N(\gamma)/\partial t=0$)
\begin{equation}
\frac{\partial}{\partial
\gamma}[(C(\gamma)-A(\gamma))N(\gamma)+D(\gamma)\frac{\partial
N(\gamma)}{\partial \gamma}]=0\;,
\label{Eq15}
\end{equation}
Chang $\&$ Cooper (1970) gave the general solution of above equation
as
\begin{equation}
N(\gamma)=x
\exp\left(-\int_{\gamma_{\rm min}}^{\gamma_{\rm max}}\frac{C(\gamma)-A(\gamma)}{D(\gamma)}d\gamma\right)\;,
\label{Eq16}
\end{equation}
where $x$ is a integration constant. Assuming $N_{\rm ini}(\gamma)$
is the initial electron distribution between $\gamma_{0min}$ and
$\gamma_{0max}$, total number of the particles in the system is
given by $N_{\rm total}=\int_{\gamma_{\rm 0min}}^{\gamma_{\rm
0max}}N_{\rm ini}(\gamma)d\gamma$, and then the integration constant
can be determined by
\begin{equation}
x=\frac{N_{\rm total}}{\exp\left(-\int_{\gamma_{\rm min}}^{\gamma_{\rm max}}\frac{C(\gamma)-A(\gamma)}{D(\gamma)}d\gamma\right)}\;,
\label{Eq17}
\end{equation}
Using
$C(\gamma)=C_{\rm 0}\gamma^{2}$ and $D(\gamma)=\gamma^{2}/2t_{\rm acc}$, we can obtain
\begin{equation}
N(\gamma)=x\gamma^{2} \exp(-2C_{\rm 0}t_{\rm acc}(\gamma-1))\;.
\label{Eq18}
\end{equation}
At the equilibrium energy $\gamma_{\rm e}$, the acceleration process
is fully compensated for by the cooling and then the acceleration
time can be given by $t_{\rm acc}=1/C_{\rm 0}\gamma_{\rm e}$. When
the electrons energy $\gamma$ equals to $\gamma_{\rm e}$, Eq. (11)
has a maximum value. In our tests, we adopt following values:
$\gamma_{\rm e}=10^{4.5}$, $N_{\rm ini}=1$ cm$^{-3}$, and
$C_{0}=3.48\times10^{-11}$ s$^{-1}$.

\begin{figure}
\epsscale{0.7} \plotone{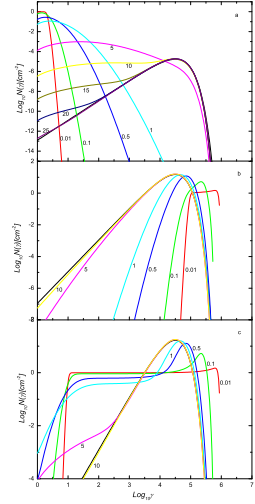} \caption{Numerical results of the
particle number density $N(\gamma,t)$ at different evolution
timescales calculated in our code for three cases. For comparison,
the analytic solutions given by Eq. (\ref{Eq18}) are shown with
black lines. The top panel(a) shows the initial electron
distribution between $\gamma_{\rm 0min}=1$ and $\gamma_{\rm 0max}=2$
(case 1), middle panel(b) shows the initial electron distribution
between $\gamma_{\rm 0min}=10^{5}$ and $\gamma_{\rm 0max}=10^{6}$
(case 2), and bottom panel(c) shows the initial electron
distribution between $\gamma_{\rm 0min}=10$ and $\gamma_{\rm
0max}=10^{6}$ (case 3). Marks near color lines represent the
evolution timescales in units of $t_{\rm acc}$.}
  \label{Fig:1}
\end{figure}

In figure \ref{Fig:1}, we show the results of the particle number
density $N(\gamma,t)$ at different evolution timescales calculated
in our numerical code for three cases. For comparison, we also show
the numerical values (black lines) in the steady state given by Eq.
(\ref{Eq18}) for three cases. In the first case, we assume an initial
electron distribution between $\gamma_{\rm 0min}=1$ and $\gamma_{\rm
0max}=2$, which means that the electron energy is much less than the
equilibrium energy $\gamma_{\rm e}$ and electrons acceleration
should dominate over all evolution processes. In this case, the
electron number density of the system decreases in the initial
energy range and simultaneously increases around the equilibrium as
the evolution timescale increases. Until evolution timescale is
increased to be more than $25t_{\rm acc}$, the system reaches the
stationary Maxwellian distribution given by Eq. (\ref{Eq18}) (see top panel
of Fig. \ref{Fig:1}). In the second case, we assume the initial
electron distribution between $\gamma_{\rm 0min}=10^{5}$ and
$\gamma_{\rm 0max}=10^{6}$, which indicates that the electron energy
is about the equilibrium energy, electrons cooling should dominate
over all evolution processes. When the evolution timescale increases
to $10t_{\rm acc}$, the system reaches the stationary Maxwellian
distribution given by Eq. (\ref{Eq18}) (see middle panel of Fig.
\ref{Fig:1}). In the third case, we assume the initial electron
distribution between $\gamma_{\rm 0min}=10$ and $\gamma_{\rm
0max}=10^{6}$.  In this case, the evolution of the electron
distribution with time depends on both electron cooling and
acceleration processes and the system reaches the stationary
Maxwellian distribution when the evolution timescale is about
$10t_{\rm acc}$ (see bottom panel of Fig. \ref{Fig:1}). In a word,
the electron spectra with larger evolution timescales ($t=20t_{\rm
acc}$ for the first case and $t\approx10t_{\rm acc}$ for the second
and third cases) calculated in our code are in agreement with the
analytic solution (i.e. Eq. (\ref{Eq18})) given by Chang $\&$ Cooper
(1970).

\section{Apply to the flaring in Markain 501}
\label{sec:apply}

\begin{figure}
\epsscale{1.0} \plotone{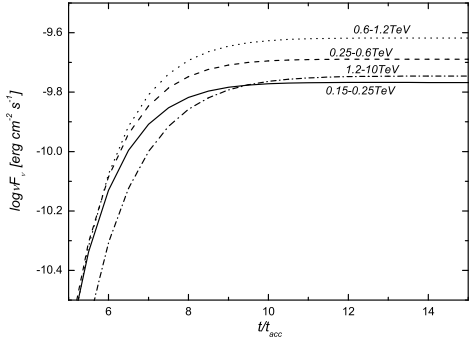}
\caption{Predicted spectra at
$0.15 - 0.25$ TeV, $0.25 - 0.6$ TeV, $0.6 - 1.2$ TeV, and $1.2 - 10$
TeV in the time-dependent one-zone SSC model. When the evolution
timescales $t\geqslant10t_{\rm acc}$, steady state spectra are
obtained.}
  \label{Fig:2}
\end{figure}

Using the time-dependent one-zone SSC solution for a spherical
geometry, we can calculate X-ray/TeV $\gamma$-ray spectra in the
stable (pre-burst) and variable (in-burst) states. In order to do
so, firstly we search for the steady state solution for electron and
photon spectra. Assuming a constant initial electron distribution
$N_{\rm ini}(\gamma,0)=2.1$ cm$^{-3}$ for
$1\leqslant\gamma\leqslant2$, we calculate the time evolution of the
spectra to $t=15t_{\rm acc}$, where the injection rate of the
electron population is $Q(\gamma)=2.1$ cm$^{-3}$ s$^{-1}$ for
$1\leqslant\gamma\leqslant2$ and a constant escape for all evolution
process are assumed. The parameters are used as follows: minimum and
maximum Lorentz factors of electrons are  $\gamma_{\rm min}=1$,
$\gamma_{\rm max}=10^{7}$, magnetic field strength is $B=0.71$ G,
emission region size is  $R=0.205\times10^{15}$ cm, Doppler factor is
$\delta=22.5$, and acceleration timescale $t_{\rm acc}=t_{\rm
esc}=R/c$. In Fig. \ref{Fig:2}, we show the changes of calculating
energy flux $\nu F_{\nu}$ at 4 energy bands of $0.15- 0.25$ TeV,
$0.25- 0.6$ TeV, $0.6- 1.2$ TeV, and $1.2- 10$ TeV with the
evolution time normalized to the acceleration time. It can be seen
from Fig. \ref{Fig:2} that the steady states for all TeV energy bands can be
reached when the evolution time $t\ge 10 t_{\rm acc}$.

\begin{figure}
\epsscale{1.0}
\plotone{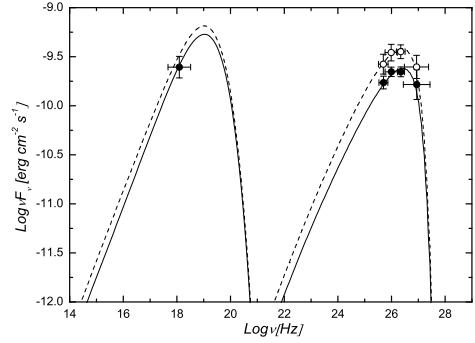}
\caption{Comparisons of predicted
multi-wavelength spectra with observed data of markarian 501 on the
July 9, 2005. solid and dash curves represent steady state (or
pre-burst) and the flaring (or in-burst) state spectra,
respectively. Observed data come from Albert et al. (2007).}
  \label{Fig:3}
\end{figure}

We assume that relativistic electrons are in the steady state during
the pre-burst of X-rays and TeV $\gamma$-rays. Therefore, we can
calculate the pre-burst X-ray/TeV $\gamma$-ray spectrum in the
one-zone SSC model using the steady state electron spectrum. In
Fig.\ref{Fig:3}, we show predicted pre-burst spectrum from X-ray to
TeV $\gamma$-ray bands (solid curve). For comparison, observed data
of markarian 501 at X-ray band and TeV band on the July 9, 2005
(Albert et al. 2007) are also shown, where black solid circles with
error bars represent the observed values at the pre-burst. It can be
seen that the observed data in the pre-burst state can be reproduced
in the SSC model.

\begin{figure}
\epsscale{1.0}
\plotone{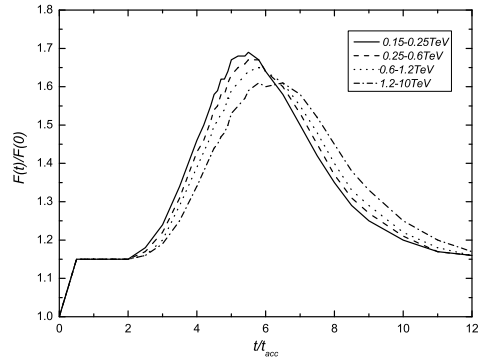}
\caption{Simulated light curves at energy bands of  0.15 - 0.25 TeV,
0.25 - 0.6 TeV, 0.6 - 1.2 TeV, and
1.2 - 10 TeV for the rapid flare of markarian 501 on July 9, 2005.
The fluxes at different wavelength are normalized to the pre-burst
state value. The quasi-symmetric shape of the light curve and
decreasing time lag of the peak with increasing energy are clearly
seen.}
  \label{Fig:4}
\end{figure}

We now consider the properties of TeV $\gamma$-ray flare of
markarian 501 in 2005 July. In order to do it, we use the physical
parameters selected above and consider the resulting steady state
spectrum as an initial condition, but we change the injection rate
of the electron population to
\begin{equation}Q(\gamma)=\left\{
\begin{array}{ll}
5.88\;\mbox{cm$^{-3}$s$^{-1}$} & \mbox{for $1\le \gamma\le 2$ and
$t\le 0.5t_{\rm acc}$}\;,\\
2.1 \;\;\mbox{cm$^{-3}$s$^{-1}$} & \mbox{otherwise}\;.
\end{array}\right.
\label{eq:Fr}
\end{equation}
Under above assumptions, we reproduce the observed TeV photon
spectrum (dash curve) of markarian 501 on the July 9, 2005 in Fig.
\ref{Fig:3}. Furthermore we simulated the light curves at energy
bands of 0.15 - 0.25 TeV, 0.25 - 0.6 TeV, 0.6 - 1.2 TeV, and 1.2 -
10 TeV respectively and show the results in Fig. \ref{Fig:4}, where
the fluxes are normalized to the pre-burst state. It can be seen
that (1) the quasi-symmetric light curve during the flare is
reproduced quite well; (2) the peaking time of the flare at higher
energies lags relative to that at lower energies; and (3) the
amplitude of the flare becomes smaller as the photon energy
increase.

\begin{figure}
\epsscale{1.0} \plotone{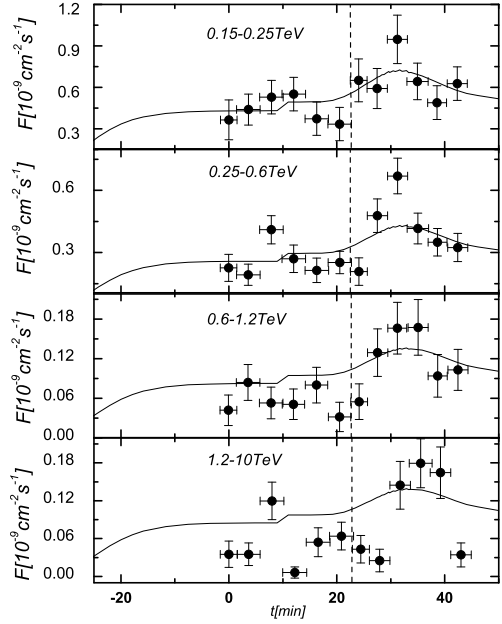} \caption{Comparisons of
simulated light curves (solid lines) with observational light curve
(data points) from the Albert et al. (2008) for the night July 9,
2005. The vertical dashed lines divide the light curves into steady
(i.e., pre-burst) and variable (i.e., in-burst) states.}
  \label{Fig:5}
\end{figure}

In order to comparing the simulated light curves with the
observations, we show the comparisons of predicted light curves with
the observations by MAGIC (Albert et al. 2008) in Fig. \ref{Fig:5}.
The integrated fluxes in the left side of the vertical dashed lines
of this figure are estimated using the differential spectra showed
in Fig. \ref{Fig:3}. It can be seen from this figure that our model
can reproduce the flare at the energy bands of 0.15 - 0.25, 0.25 -0.6,
0.6 - 1.2, and 1.2-10 TeV.

\begin{figure}
\epsscale{1.0} \plotone{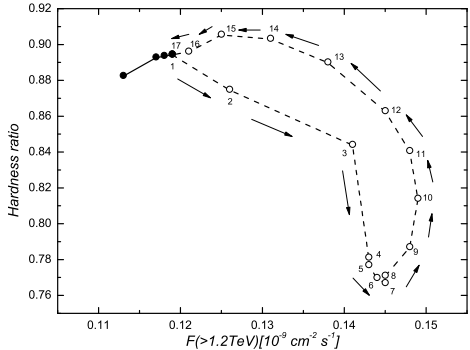} \caption{Simulated the hardness
ratio $\frac{F(1.2 - 10\rm TeV)}{F(0.6 - 1.2\rm TeV)}$ versus flux
$F(>1.2~\rm TeV)$ for the night July 9, 2005. Black and open circles
represent pre-burst  and in-burst emission respectively. The number
near the markers denote the position of the points in the light
curves. A canonical anti-clockwise-loop is seen.}
  \label{Fig:6}
\end{figure}

Finally, we calculate the time lag between 0.15 - 0.25 TeV and 1.2 -
10 TeV using the Gaussian fit of the simulated light curves, and
find out that the light curve at higher energies (1.2 - 10 TeV) lags
relative to that at lower energies (0.15 - 0.25 TeV) by a factor of
about $0.9t_{\rm acc}$, this timescale correspond to about 4.7
minutes in the observer's frame. We also calculate the evolution of
the hardness ratio which is defined as the ratio $\frac{F(1.2 -
10~\rm TeV)}{F(0.6 - 1.2~\rm TeV)}$. The evolution of the hardness
ratio with the emitted flux above $1.2$ TeV is shown in Fig. \ref{Fig:6}. It
can be seen from Fig. \ref{Fig:6} that the evolution of the flare points shows
a clear canonical anti-clockwise-loop.

\section{Discussion and conclusions}
\label{sec:discussion}

In this paper, we have tried to explain the TeV $\gamma$-ray flare
of markarian 501 observed by MAGIC telescope on July 9, 2005 in the
context of the time-dependent one-zone SSC model which includes
stochastic particle acceleration. In this model, particles with low
energy are assumed to be injected and then are accelerated to higher
energy by second-order Fermi acceleration mechanism (Fermi 1949),
the most important photon targets for inverse Compton scattering by
relativistic electrons are the synchrotron photons. We have studied
time-dependent properties of flares by reproducing the pre-burst
spectrum of the source and varying the injection rate. In our
results, the behaviour of the mean multi-frequencies spectra before
and during the flare is a little different, the peaks of both
synchrotron and IC emissions move to lower frequencies, we argue
that this can be explained by the energy loss of the electrons
during the outburst. In this scenario, the hard lag flare can be
obtained and during the flare, it shows a clear canonical
anti-clockwise-loop.

It should be noted that hard lags require some sort of particle
acceleration. If the variability timescale is faster than the
cooling timescale, the radiation from accelerated particles would
show a hard lag (Albert et al. 2008). Kirk et al. (1998) argued that
the hard lag from the acceleration process induces to the
anti-clockwise-loop pattern. In this view, Albert et al. (2007)
concluded that the acceleration process of low energy particles
probably dominate over the TeV $\gamma$-ray flare. Assuming a low
energy electrons injection and stochastic acceleration, our
calculations predicted hard lags dependent flaring activity and
showed a anti-clock-loop evolution of the hardness ratio with the
flux. These are in agreement with the observational results on July
9, 2005, and imply that, during the flare, the dynamics of the
system is dominated by the acceleration, rather than by the cooling
processes. However, a detailed investigation of electron
acceleration in the presence of losses has so far been performed
only by a few investigators (e.g. Mastichiadis \& Moraits 2008).
Given the complexity of the flaring activity of high energy
radiation, this requires more detailed observations and the issue to
be open.

The magnetohydrodynamic turbulence will is generated if standing
shocks form in the neighborhood of the central object, which amplify
any incoming upstream turbulence in the downstream accretion shock
magnetosheath (Campeanu \& Schlickeiser 1992). These
magnetohydrodynamic plasma waves are the free energy and lead to
stochastic acceleration of charged particles. Actually, stochastic
acceleration occurs wherever there are turbulent magnetic fields and
can spread to an extended region, the size is determined by the
turbulence generation and decay rates. Virtanen \& Vainio (2005)
simulated the stochastic acceleration in relativistic shocks and
shows, when the particles were accelerated behind the discontinuity,
a gradual shift of the whole particle spectrum to higher energy.
Some recent observations of particles spectra with hard power law
spectral indices, $N(\gamma)\propto\gamma^{-n}$ with $n <2$,
suggests that the stochastic acceleration is seen in the
observations (Katarzynski et al. 2006; B$\ddot{o}$ttcher et al.
2008). The model presented here contains the stochastic acceleration
process. For simplicity, we introduced a constant acceleration term,
which is associated with the momentum diffusion coefficient
$D(p,t)$. The form of the diffusion coefficient due to interactions
with magnetohydrodynamic waves has been discussed in details (e.g.
Kulsrud \& Ferrari 1971; Schlickeiser 2002). In our model, both
constant acceleration and escape times are assumed, leading to
$D(\gamma,t)=\gamma^{2}/2t_{\rm acc}\propto\gamma^{2}$. The form of
the diffusion coefficient corresponds to the hard-sphere
approximation, in which the mean free path for article-wave
interaction is independent of particle energy, and probably induces
to a complicated spectrum. Furthermore, since the basic shock
acceleration models postulate that $t_{\rm acc}\simeq t_{\rm esc}$
(e.g., Katarzynski et al. 2006), we adopt the shorter acceleration
and escape timescales ($t_{\rm acc}=t_{\rm esc}=\frac{R}{c}=t_{\rm
cr}$) than other investigators (generally, $t_{\rm acc}>t_{\rm cr}$,
and $t_{\rm esc}>t_{\rm cr}$, see e.g., Kirk et al. 1998;
Mastichiadis \& Moraitis 2008). These assumptions can lead to higher
acceleration rate and lower escape rate, and make more particles
acceleration up to high energy rapidly.

There are two scenarios for explaining the intrinsic variability.
The first scenario assumes that the observed variations origin from
the geometry of emitting sources (e.g., Camenzind \& Krockenberger
1992; Gopal-Krishna \& Wiita 1992). The second scenario assumes that
the variability is generated by change of the emission condition. A
typical example is that fresh particles are injected into
acceleration region and then are accelerated (e.g., Blandford \&
Konigl 1979; Marscher \& Gear 1985; Celotti et al. 1991; Kirk et al.
1998). In order to reproduce both high energy radiation and
variability of markarian 501, we change the injection rate of the
low energy particles. It should be noted that when the shock front
overruns a region in the jet in which the local plasma density is
enhanced. The number of particles increase as an avalanche occurring
in the jet, the injection rate can be expected to change.

\section*{Acknowledgments}
We thank the anonymous referee for valuable comments and suggestions.
This work is partially supported by the National Natural Science
Foundation of China under grants 10763002 and 10778702 and the
Natural Science Foundation of Yunnan Province under grants
2009ZC056M, 2008CC011. This work is also supported by the Science
Foundation of Yunnan educational department (grant 08Z0020).


\end{document}